\documentclass[11pt,letterpaper]{amsart}

\usepackage[letterpaper,top=3cm, bottom=3cm, left=3.5cm, right=3.5cm]{geometry}

\makeatletter
\def\subsection{\@startsection{subsection}{2}%
  \z@{.7\linespacing\@plus.7\linespacing}{.2\linespacing}%
  {\centering\normalfont\scshape}}
\makeatother

\usepackage{amssymb}
\usepackage{enumerate}

\newtheorem{theorem}{Theorem}[subsection]
\newtheorem{proposition}[theorem]{Proposition}
\newtheorem{corollary}[theorem]{Corollary}
\newtheorem{lemma}[theorem]{Lemma}
\newtheorem{construction}[theorem]{Construction}
\newtheorem{conjecture}[theorem]{Conjecture}
\newtheorem{question}[theorem]{Question}
\theoremstyle{definition}
\newtheorem{definition}[theorem]{Definition}
\newtheorem{example}[theorem]{Example}

\numberwithin{equation}{section}

\newcommand{\norm}[1]{\lVert#1\rVert}
\newcommand{\abs}[1]{\lvert#1\rvert}
\newcommand{\bigabs}[1]{\big\lvert#1\big\rvert}

\newcommand{\biggabs}[1]{\bigg\lvert#1\bigg\rvert}
\newcommand{\Biggabs}[1]{\Bigg\lvert#1\Bigg\rvert}

\newcommand{\e}{\epsilon}

\DeclareMathOperator{\Tr}{Tr}

\DeclareMathOperator{\E}{E}
\DeclareMathOperator{\sign}{sign}
\DeclareMathOperator{\lcm}{lcm}

\DeclareMathOperator{\ord}{ord}

\newcommand{\C}{\mathbb{C}}
\newcommand{\F}{\mathbb{F}}
\newcommand{\N}{\mathbb{N}}
\newcommand{\R}{\mathbb{R}}
\newcommand{\Z}{\mathbb{Z}}

\newcommand{\Bf}{\mathfrak{B}}
\newcommand{\Df}{\mathfrak{D}}
\newcommand{\Fc}{\mathcal{F}}

\begin{document}

\title{Sequences with small correlation}

\author{Kai-Uwe Schmidt}

\address{Department of Mathematics, Paderborn University, Warburger Str.\ 100, 33098 Paderborn, Germany}
\email{kus@math.upb.de}

\date{06 October 2015}


\begin{abstract}
The extent to which a sequence of finite length differs from a shifted version of itself is measured by its aperiodic autocorrelations. Of particular interest are sequences whose entries are $1$ or $-1$, called binary sequences, and sequences whose entries are complex numbers of unit magnitude, called unimodular sequences. Since the 1950s, there is sustained interest in sequences with small aperiodic autocorrelations relative to the sequence length. One of the main motivations is that a sequence with small aperiodic autocorrelations is intrinsically suited for the separation of signals from noise, and therefore has natural applications in digital communications. This survey reviews the state of knowledge concerning the two central problems in this area: How small can the aperiodic autocorrelations of a binary or a unimodular sequence collectively be and how can we efficiently find the best such sequences? Since the analysis and construction of sequences with small aperiodic autocorrelations is closely tied 
to the (often much easier) analysis of periodic autocorrelation properties, several fundamental results on corresponding problems in the periodic setting are also reviewed.
\end{abstract}

\maketitle


\section{Introduction}

By a \emph{sequence} of length $n$ we mean an element of $\C^n$. For a sequence $A$ of length~$n$, we denote by $A(k)$ the $k$-th entry in $A$ (starting with $k=0$). It is convenient to allow~$k$ to be an arbitrary integer and reduce $k$ modulo~$n$ if necessary. It is desirable from a practical viewpoint and appealing from a theoretical viewpoint to restrict the entries of a sequence to a small set. The most interesting case occurs when the entries are just $-1$ or $1$, in which case we call the sequence \emph{binary}. 
\par
Let $A$ be a sequence of length $n$. For an integer $u$ with $0\le u<n$, let
\[
C_u(A)=\sum_{0\le k,k+u<n}A(k)\overline{A(k+u)}
\]
be the \emph{aperiodic autocorrelation} of $A$ at shift $u$. We call $C_0(A)$, the sum of squared magnitudes of entries of $A$, the \emph{trivial} aperiodic autocorrelation of $A$ and the values of~$C_u(A)$ for all nonzero $u$ the \emph{nontrivial} aperiodic autocorrelations of $A$.
\par
There is sustained interest in sequences with restricted entries such that their nontrivial aperiodic autocorrelations are small with respect to some measure. For example, Turyn~\cite{Tur1960} asked for binary sequences having the ideal property that all nontrivial aperiodic autocorrelations are in the set $\{-1,0,1\}$. Such sequences are now called \emph{Barker} sequences, since a related problem was studied earlier by Barker~\cite{Bar1953}. The problem as to whether there exist infinitely many Barker sequences is still open, although there is overwhelming evidence that there is no Barker sequence of length greater than $13$. Many of the problems discussed in this survey are motivated by the apparent nonexistence of long Barker sequences. The most natural question is to ask for binary sequences for which the magnitudes of the nontrivial aperiodic autocorrelations are collectively as small as possible. This problem will be discussed in Section~\ref{sec:aperiodic}.
\par
For a sequence $A$ of length $n$ and an integer $u$, let 
\[
R_u(A)=\sum_{k=0}^{n-1}A(k)\overline{A(k+u)}
\]
be the \emph{periodic autocorrelation} of $A$ at shift $u$. Again, we call $R_0(A)$ the \emph{trivial} periodic autocorrelation of $A$ and the values of $R_u(A)$ for all nonzero $u$ the \emph{nontrivial} periodic autocorrelations of $A$. The relationship between the aperiodic and periodic autocorrelations of a sequence $A$ of length $n$ is given by
\begin{equation}
R_u(A)=C_u(A)+\overline{C_{n-u}(A)}\qquad\text{for $0<u<n$}.   \label{eqn:R_from_C}
\end{equation}
The periodic autocorrelations are usually much easier to study than their aperiodic counterparts. Indeed, a typical attempt to obtain sequences with good aperiodic autocorrelations is to identify sequences with good periodic autocorrelations and then examine their aperiodic autocorrelations. We shall see that this approach often works well. Periodic autocorrelations also arise in settling the existence question for Barker sequences since a putative Barker sequence of length greater than $13$ must have all of its nontrivial periodic autocorrelations equal to zero; such a sequence is called \emph{perfect}. The existence of infinitely many perfect binary sequences is also still unsettled, although likewise there is overwhelming evidence that there is no perfect binary sequence of length greater than $4$. While the study of periodic autocorrelations is historically at least partly motivated by questions involving aperiodic autocorrelations, many challenging problems have since arisen in the periodic case and the 
field has become a very active research area. Some fundamental topics will be discussed in Section~\ref{sec:periodic}.
\par
The apparent nonexistence of long Barker sequences has led researchers to study alternative objects by relaxing various constraints. One possibility, discussed in Section~\ref{sec:nonbinary}, is to consider \emph{$H$-phase} sequences, namely sequences whose entries are $H$-th roots of unity, and \emph{unimodular} sequences, namely sequences whose entries have unit magnitude. Another possibility, discussed in Section~\ref{sec:Golay}, is to consider \emph{Golay pairs}, namely pairs of sequences whose aperiodic autocorrelations sum to zero for each nonzero shift. 
\par
There are several other works that survey topics involving correlations of sequences. Like the present survey, most of them focus on particular aspects. Some recommended articles that also helped me in preparing the present survey are: Turyn~\cite{Tur1968}, which covers the essential knowledge until 1968; Jungnickel and Pott~\cite{JunPot1999} and Cai and Ding~\cite{CaiDin2009}, which concentrate on optimal binary sequences and cyclic difference sets; Helleseth and Kumar~\cite{HelKum1998} and Golomb and Gong~\cite{GolGon2005}, whose focus is on periodic correlations; Jedwab~\cite{Jed2008}, whose focus is on aperiodic autocorrelations; and Jedwab~\cite{Jed2005} and H{\o}holdt~\cite{Hoh2006}, which survey results on the merit factor problem for binary sequences until 2006. For the interested reader, I also recommend Borwein~\cite{Bor2002}, which covers some material of this survey and exhibits many interesting connections to analysis and number theory.


\section{Periodic autocorrelation of binary sequences}
\label{sec:periodic}

\subsection{Bounds and constructions}
\label{sec:bounds_and_constructions}

In this section we are interested in binary sequences for which the nontrivial periodic autocorrelations are as small as possible in magnitude. From this viewpoint, an ideal binary sequence has all nontrivial periodic autocorrelations equal to zero. Such a sequence is called \emph{perfect}.  However, the only length $n>1$ for which a perfect sequence is known is $n=4$. For example, $(+++\,-)$ is a perfect sequence (writing $+$ for $1$ and~$-$ for $-1$). In Section~\ref{sec:Nonexistence_perfect} we shall discuss some results establishing the nonexistence of perfect sequences. 
\par
A simple necessary condition for the existence of a perfect binary sequence is contained in the following lemma, which follows from a simple parity argument.
\begin{lemma}
\label{lem:Ru_mod_four}
All periodic autocorrelations of a binary sequence of length $n$ are congruent to $n$ modulo $4$.
\end{lemma}
\par
Lemma~\ref{lem:Ru_mod_four} implies that every binary sequence $A$ of length $n>1$ satisfies
\begin{equation}
\max_{0<u<n}\abs{R_u(A)}\ge\begin{cases}
0 & \text{for $n\equiv 0\pmod 4$}\\
1 & \text{for $n\equiv 1$ or $3\pmod 4$}\\
2 & \text{for $n\equiv 2\pmod 4$}
\end{cases}   \label{eqn:Ru_lower_bound}
\end{equation}
and so perfect binary sequences can exist only when the length is divisible by $4$. Indeed since 
\begin{equation}
\sum_{u=0}^{n-1}R_u(A)=\Bigg|\sum_{k=0}^{n-1}A(k)\Bigg|^2,   \label{eqn:sum_of_Ru}
\end{equation}
the length of a perfect binary sequence must be an even square. We call a binary sequence $A$ \emph{optimal} if equality holds in~\eqref{eqn:Ru_lower_bound}. We shall see below that there are infinitely many lengths congruent to $2$ or $3$ modulo $4$ for which optimal binary sequences exist. However, if $n\equiv 1\pmod 4$, then optimal binary sequences are known only for $n=5$ or~$13$. For example,
\begin{equation}
(+++-+)\quad\text{and}\quad (+++++--++-+-+)   \label{eqn:optimal_1_mod_4}
\end{equation}
are optimal binary sequences of length $5$ and $13$, respectively. Some nonexistence results will be discussed in Section~\ref{sec:Nonexistence_perfect}.
\par
Sometimes, applications require \emph{balanced} binary sequences, by which we mean binary sequences $A$ of length $n$ satisfying
\[
\Biggabs{\sum_{k=0}^{n-1}A(k)}\le 1.
\]
It follows from Lemma~\ref{lem:Ru_mod_four} and the identity~\eqref{eqn:sum_of_Ru} that an optimal binary sequence $A$ of length $n$ cannot be balanced if $n$ is congruent to $0$ or $1$ modulo $4$. Therefore, every balanced binary sequence $A$ of length $n>1$ satisfies
\begin{equation}
\max_{0<u<n}\abs{R_u(A)}\ge\begin{cases}
1 & \text{for $n\equiv 3\pmod 4$}\\
2 & \text{for $n\equiv 2\pmod 4$}\\
3 & \text{for $n\equiv 1\pmod 4$}\\
4 & \text{for $n\equiv 0\pmod 4$}.
\end{cases}   \label{eqn:Ru_lower_bound_balanced}
\end{equation}
If $A$ is a balanced binary sequence of length $n>1$ for which equality holds in~\eqref{eqn:Ru_lower_bound_balanced}, then we say that $A$ is \emph{optimal balanced}.
\par
We now show that optimal balanced binary sequences exist for infinitely many lengths of every congruence class modulo $4$. 
\begin{definition}[Legendre sequences]
For an odd prime $p$, a \emph{Legendre sequence}~$A$ of length $p$ is defined by
\[
A(k)=\begin{cases}
1  & \text{for $p\mid k$ or $k$ a square modulo $p$}\\
-1 & \text{otherwise}.
\end{cases}
\]
\end{definition}
The following result is classical (see~\cite{Pal1933}, for example).
\begin{theorem}
\label{thm:Legendre_periodic}
Legendre sequences are optimal balanced. In particular, the nontrivial periodic autocorrelations of a Legendre sequence of length $p$ are equal to $-1$ if $p\equiv 3\pmod 4$ and are in the set $\{1,-3\}$ if $p\equiv 1\pmod 4$.
\end{theorem}
\par
Therefore there exist optimal balanced binary sequences for all odd prime lengths. To obtain optimal balanced binary sequences of even length, we require the following definition.
\begin{definition}[Sidelnikov sequences]
Let $q$ be an odd prime power and let $\theta$ be a primitive element of $\F_q$. Define a sequence $A$ of length $q-1$ by
\[
A(k)=\begin{cases}
1  & \text{if $\theta^k+1$ is zero or a square in $\F_q$}\\
-1 & \text{otherwise}.
\end{cases}
\]
\end{definition}
It is customary to call the above defined sequences \emph{Sidelnikov sequences}. However, to my knowledge, they were first considered by Turyn~\cite[p.~208-209]{Tur1968} and were later studied independently by Sidelnikov~\cite{Sid1969} and Lempel, Cohn, and Eastman~\cite{LemCohEas1977}.
\begin{theorem}[{\cite{Sid1969},~\cite{LemCohEas1977}}]
Sidelnikov sequences are optimal balanced. In particular, the nontrivial periodic autocorrelations of a Sidelnikov sequence of length $n$ are in the set $\{-2,2\}$ if $n\equiv 2\pmod 4$ and are in the set $\{0,-4\}$ if $n\equiv 0\pmod 4$.
\end{theorem}
\par
Several other constructions of optimal balanced binary sequences are known, as surveyed in detail by Cai and Ding~\cite{CaiDin2009}. The currently known constructions in the case that~$n$ is congruent to $3$ modulo $4$ will be reviewed in Section~\ref{sec:difference_sets}. For now, we consider one more important class of binary sequences, namely the \emph{Galois sequences}, which are also known as \emph{m-sequences}. Recall that the \emph{absolute trace function} on $\F_{2^m}$ is the mapping $\Tr:\F_{2^m}\to\F_2$ given by
\[
\Tr(y)=\sum_{j=0}^{m-1}y^{2^j}.
\]
\begin{definition}[Galois sequences]
Let $\theta$ be a primitive element of $\F_{2^m}$ and let $a\in\F_{2^m}$ be nonzero. A \emph{Galois sequence} $A$ of length $2^m-1$ is defined by
\[
A(k)=\begin{cases}
1  & \text{for $\Tr(a\theta^k)=0$}\\
-1 & \text{for $\Tr(a\theta^k)=1$}.
\end{cases}
\]
\end{definition}
\par
Note that the cyclic shifts of a Galois sequence are also Galois sequences. Galois sequences can be equivalently defined (and efficiently generated) using linear feedback shift registers~\cite{Golomb1967}.
\par
The following result is an immediate consequence of elementary properties of the trace function.
\begin{theorem}
Galois sequences are optimal balanced. In particular, the nontrivial periodic autocorrelations of a Galois sequence are equal to $-1$.
\end{theorem}
\par
In fact, Galois sequences have a stronger property than balancedness. If $A$ is a Galois sequence of length $2^m-1$ and $k$ takes on all values in the set $\{0,1,\dots,2^m-2\}$, then the $m$-tuples
\[
(A(k),A(k+1),\dots,A(k+m-1))
\]
range through all $2^m-1$ possible binary sequences of length $m$, except for the all-ones sequence (see~\cite{Golomb1967} or~\cite{GolGon2005}, for example).

\subsection{Cyclic difference sets}
\label{sec:difference_sets}

In this section we consider binary sequences whose nontrivial periodic autocorrelations are all equal, say to $\gamma$. Such sequences are said to possess a \emph{two-level} periodic autocorrelation (with one level being the trivial periodic autocorrelation) and are equivalent to cyclic difference sets.
\par
A \emph{difference set} with parameters $(n,k,\lambda)$ is a $k$-subset $D$ of a finite group $G$ of order~$n$ such that every non-identity element $g$ of $G$ has exactly $\lambda$ representations $g=xy^{-1}$ for $x,y\in G$ (so that $k(k-1)=\lambda(n-1)$). If $G$ is a cyclic group, then we say that the difference set is \emph{cyclic}. Note that the complement of a difference set is also a difference set, so we may assume that $k\le n/2$.
\par
Let $G$ be a cyclic group of order $n$ and fix a generator $\omega$ of $G$. Given a subset $D$ of~$G$, we associate with $D$ a binary sequence $A$ of length $n$ via
\[
A(k)=\begin{cases}
-1 & \text{for $\omega^k\in D$}\\
1  & \text{for $\omega^k\not\in D$}.
\end{cases}
\]
We call $A$ the \emph{characteristic sequence} of $D$ (with respect to $\omega$). The following result is classical and readily verified.
\begin{proposition}
Let $D$ be a subset of a cyclic group. Then the characteristic sequence of $D$ has two-level periodic autocorrelation if and only if $D$ is a difference set. Moreover, if $D$ is a difference set with parameters $(n,k,\lambda)$, then the nontrivial periodic autocorrelations of its characteristic sequence equal $n-4(k-\lambda)$.
\end{proposition}
\par
In the case $n\not\equiv 2\pmod 4$, an optimal binary sequence is equivalent to a cyclic difference set with parameters
\begin{alignat}{3}
&\bigg(n,\frac{n-\sqrt{n}}{2},\frac{n-2\sqrt{n}}{4}\bigg)&&\quad\text{for $n\equiv 0\pmod 4$},   \label{eqn:parameters_n0}\\
&\bigg(n,\frac{n-\sqrt{2n-1}}{2},\frac{n+1-2\sqrt{2n-1}}{4}\bigg)&&\quad\text{for $n\equiv 1\pmod 4$},   \label{eqn:parameters_n1}\\
&\bigg(n,\frac{n-1}{2},\frac{n-3}{4}\bigg)&&\quad\text{for $n\equiv 3\pmod 4$}.   \label{eqn:parameters_n3}
\end{alignat}
\par
As mentioned previously, there are only finitely many known cyclic difference sets with parameters~\eqref{eqn:parameters_n0} or~\eqref{eqn:parameters_n1}. All known cyclic difference sets with parameters~\eqref{eqn:parameters_n3} occur when $n$ is either a prime number, a product of twin primes, or a Mersenne number. Examples are given by Legendre sequences of length~$p$ satisfying $p\equiv 3\pmod 4$ and by Galois sequences, in which cases the sets are called \emph{Paley} and \emph{Singer} difference sets, respectively, since related structures were first studied by Paley~\cite{Pal1933} and Singer~\cite{Sin1938}. There are several other constructions of such difference sets, or equivalently optimal binary sequences of length $n\equiv 3\pmod 4$, which we shall review briefly.

\subsubsection*{The twin-prime construction~\cite{Bra1953}}
Let $p$ and $p+2$ be prime numbers and let $X$ and $Y$ be Legendre sequences of length $p$ and $p+2$, respectively. The sequence $A$ of length $p(p+2)$ given by
\[
A(k)=\begin{cases}
X(k)Y(k) & \text{for $p\nmid k$ and $p+2\nmid k$}\\
1        & \text{for $p\mid k$ and $p+2\nmid k$}\\
-1       & \text{for $p+2\mid k$}
\end{cases}
\]
is the characteristic sequence of a difference set with parameters~\eqref{eqn:parameters_n3}.

\subsubsection*{The Hall construction~\cite{Hal1956}}

Let $p$ be a prime number of the form $4x^2+27$ for $x\in\Z$ and let $\theta$ be a primitive root modulo $p$. Let $C_k$ be the set of numbers $a\in\Z$ for which the congruence $x^6\theta^k\equiv a\pmod p$ has a solution $x\in\Z$. Let $D$ be either $C_0\cup C_1\cup C_3$ or $C_0\cup C_3\cup C_5$, depending on whether $3$ is contained in $C_1$ or $C_5$, respectively. (By quadratic and cubic reciprocity laws we always have $3\in C_1\cup C_5$.) The sequence $A$ of length $p$ given by
\[
A(k)=\begin{cases}
-1 & \text{for $k\in D$}\\
 1 & \text{otherwise}
\end{cases}
\]
is the characteristic sequence of a difference set with parameters~\eqref{eqn:parameters_n3}, called a \emph{Hall difference set}.

\subsubsection*{The Maschietti construction~\cite{Mas1998}}

Let $\theta$ be a primitive element of $\F_{2^m}$ and let $t$ be an integer coprime to $2^m-1$ such that the mapping from $\F_{2^m}$ to $\F_{2^m}$, given by $x\mapsto x^t+x$, is $2$-to-$1$. The sequence $A$ of length $2^m-1$ given by
\[
A(k)=\begin{cases}
-1 & \text{if $y^t+y=\theta^k$ has a solution $y\in\F_{2^m}$}\\
 1 & \text{otherwise}
\end{cases}
\]
is the characteristic sequence of a difference set with parameters~\eqref{eqn:parameters_n3}. This construction was first given by Maschietti~\cite{Mas1998} by establishing a link to monomial hyperovals in finite projective planes. The above description follows Evans, Hollmann, Krattenthaler, and Xiang~\cite{EvaHolKraXia1998}. Up to equivalences, the only known choices for $t$ are $t=2^i$ for $\gcd(i,m)=1$ (in which case we obtain Galois sequences again), $t=6$ for odd $m$, $t=3\cdot 2^{(m+1)/2}+4$ for odd $m$, $t=2^{(m+1)/2}+2^{(3m+1)/4}$ for $m\equiv 1\pmod 4$, and $t=2^{(m+1)/2}+2^{(m+1)/4}$ for $m\equiv 3\pmod 4$.

\subsubsection*{The Dillon-Dobbertin construction~\cite{DilDob2004}}

Let $\theta$ be a primitive element of $\F_{2^m}$, let $t$ be an integer coprime to $m$ satisfying $0<t<m/2$, and write $d=4^t-2^t+1$. The sequence~$A$ of length $2^m-1$ given by
\[
A(k)=\begin{cases}
-1 & \text{if $(y+1)^d+y^d+1=\theta^k$ has a solution $y\in\F_{2^m}$}\\
 1 & \text{otherwise}
\end{cases}
\]
is the characteristic sequence of a difference set with parameters~\eqref{eqn:parameters_n3}.

\subsubsection*{The No-Chung-Yun construction~\cite{DilDob2004}}

Let $m$ be an integer that is not divisible by $3$, write $t=(m\pm 1)/3$ depending on the congruence class of $m$ modulo $3$ (so that $t$ is integral), and write $d=4^t-2^t+1$. The sequence~$A$ of length $2^m-1$ given by
\[
A(k)=\begin{cases}
-1 & \text{if $(y+1)^d+y^d=\theta^k$ has a solution $y\in\F_{2^m}$}\\
 1 & \text{otherwise}
\end{cases}
\]
is the characteristic sequence of a difference set. For even $m$, this difference set has parameters~\eqref{eqn:parameters_n3}. For odd $m$, its complement has parameters~\eqref{eqn:parameters_n3}. This construction was given by No, Chung, and Yun~\cite{NoChuYun1998} and the autocorrelation properties were proved by Dillon and Dobbertin~\cite{DilDob2004}.

\subsubsection*{The Gordon-Mills-Welch construction~\cite{GorMilWel1962}}

This construction produces new cyclic difference sets from known ones. Let $s$ and $m$ be integers with $1<s<m$ and $s\mid m$. Let~$D$ be a difference set in $\F_{2^s}^*$ with parameters $(2^s-1,2^{s-1},2^{s-2})$ (so that its complement has parameters~\eqref{eqn:parameters_n3}). Let $C$ be the set of elements $c\in \F_{2^m}$ with $\Tr_{2^m/2^s}(c)=1$, where $\Tr_{2^m/2^s}$ is the \emph{relative trace} from $\F_{2^m}$ to $\F_{2^s}$, given by
\[
\Tr_{2^m/2^s}(y)=\sum_{j=0}^{m/s-1}y^{2^{sj}}.
\]
Then $\{cd:c\in C,d\in D\}$ is a difference set in $\F_{2^m}^*$ with parameters $(2^m-1,2^{m-1},2^{m-2})$. 
\par
This construction is a rich source of cyclic difference sets because we do not require any further information about $D$. In particular, the construction can be iterated. If $D$ is a Singer difference set, then the characteristic sequence of the new difference set is sometimes called a \emph{GMW sequence}~\cite{SchWel1984}.


\subsection{Nonexistence results}
\label{sec:Nonexistence_perfect}

We have seen that there are infinite families of optimal binary sequences whose lengths are congruent to either $2$ or $3$ modulo $4$. In this section, we review nonexistence results for optimal binary sequences whose lengths are congruent to either $0$ or $1$ modulo~$4$. In these cases the sequences are in one-to-one correspondence with cyclic difference sets. It is customary and convenient to identify the parameters of an $(n,k,\lambda)$ difference set with the tuple $(n,k,\lambda,m)$, where $m=k-\lambda$.
\par
Our main focus is on the case $n\equiv 0\pmod 4$, in which case the difference sets have parameters
\[
(4u^2,2u^2-u,u^2-u,u^2)
\]
and are called \emph{cyclic Hadamard difference sets}. Some comments on the case $n\equiv 1\pmod 4$ will be given at the end of this section.
\par
There is a well known relationship between perfect binary sequences and Hadamard matrices; a square matrix $H$ of order $n$ is a \emph{Hadamard matrix} if all of its entries are $-1$ or $1$ and $HH^T=nI$, where $I$ is the $n\times n$ identity matrix. It is readily verified that the circulant matrix of order $n$ corresponding to a binary sequence of length $n$ is a Hadamard matrix if and only if the sequence is perfect. For example, writing $+$ for~$1$ and $-$ for $-1$, the perfect binary sequence $(+++\,-)$ gives the circulant Hadamard matrix
\[
\begin{pmatrix}
+ & + & + & -\\
- & + & + & +\\
+ & - & + & +\\
+ & + & - & +
\end{pmatrix}.
\]
An old conjecture due to Ryser~\cite[p.~134]{Rys1963} asserts that there are no circulant Hadamard matrices of order greater than $4$. Equivalently, we have the following.
\begin{conjecture}[\cite{Rys1963}]
\label{con:perfect_binary_sequences}
There is no perfect binary sequence of length $n>4$.
\end{conjecture}
\par
This conjecture is still open. However strong partial results are known and the most important methods will be reviewed below. It should be emphasised that most of these methods can be applied to difference sets that are not necessarily cyclic and sometimes even to more general combinatorial objects. However, we will restrict ourselves to the case of cyclic difference sets.
\par
We have seen that the length of a perfect binary sequence must be an even square. Turyn~\cite[p.~336]{Tur1965} proved the much deeper result that the length must actually be $4$ times an odd square.
\begin{theorem}[\cite{Tur1965}]
\label{thm:Hadamard_diff_set_u_odd}
If there exists a perfect binary sequence of length $n\ge 4$, then $n=4u^2$ for an odd integer $u$.
\end{theorem}
\par
We proceed with a classical result due to Turyn~\cite{Tur1965}, for which we require the following definition. For integers $a$ and $w>0$, we say that~$a$ is \emph{semiprimitive} modulo $w$ if there exists an integer $t$ such that $a^t\equiv -1\pmod w$ and we say that $a$ is \emph{self-conjugate} modulo $w$ if each prime divisor $p$ of $a$ is semiprimitive modulo $w_p$, where $w_p$ is the largest divisor of~$w$ that is not divisible by $p$.
\par
The following result is \cite[Corollary 1]{Tur1965} specialised to cyclic difference sets.
\begin{theorem}[\cite{Tur1965}]
\label{thm:self_conjugacy_bound}
Suppose that there exists a cyclic difference set with parameters $(n,k,\lambda,m)$. Suppose further that there are positive integers $c$ and $d$ satisfying $\gcd(c,d)>1$ such that $d\mid n$ and $c^2\mid m$ and such that $c$ is self-conjugate modulo $d$. Let~$r$ be the number of distinct prime divisors of $\gcd(c,d)$. Then $cd\le 2^{r-1}n$.
\end{theorem}
\par
For convenience, we state Theorem~\ref{thm:self_conjugacy_bound} for perfect binary sequences.
\begin{corollary}
\label{cor:self_conjugacy_bound}
Suppose that there exists a perfect binary sequence of length $n=4u^2$. Suppose further that there are positive integers $c$ and $d$ satisfying $\gcd(c,d)>1$ such that $d\mid n$ and $c\mid u$ and such that $c$ is self-conjugate modulo $d$. Let $r$ be the number of distinct prime divisors of $\gcd(c,d)$. Then $cd\le 2^{r-1}n$.
\end{corollary}
\par
Corollary~\ref{cor:self_conjugacy_bound} is particularly useful if $u$ has a relatively large odd prime factor. Indeed, if~$p$ is an odd prime such that $u=p^av$ for positive integers $a$ and $v$, then take $c=p^a$ and $d=2p^{2a}$ in Corollary~\ref{cor:self_conjugacy_bound} to conclude that no perfect binary sequence of length $4u^2$ exists if $p^a>2v^2$. In particular, taking $v=1$, we see that there is no perfect binary sequence whose length is four times an odd prime power.
\par
Corollary~\ref{cor:self_conjugacy_bound} proves the nonexistence of perfect binary sequences of length $4u^2$ for all~$u$ satisfying $1<u<55$, except for $u=39$. However, this last case was also ruled out by Turyn~\cite[p.~202]{Tur1968}.
\par
It took more than thirty years until the next open case $u=55$ was disqualified by B.~Schmidt \cite{Sch1999},~\cite{Sch2002} with the invention of a powerful method, known as the ``Field Descent Method''. This method was subsequently refined by Leung and B.~Schmidt \cite{LeuSch2005}, \cite{LeuSch2011}. The results involve a rather technical function $F(n,m)$, which we define below (our definition is taken from~\cite{LeuSch2005} and is equivalent to the original definition of~\cite{Sch1999} modulo a slight inaccuracy). Recall the following standard notation. For integers $a$ and $w>0$, the number $\ord_w(a)$ is the smallest positive integer~$t$ such that $a^t\equiv 1\pmod w$. For positive integers $r$ and $b$, the number $\nu_r(b)$ is the largest integer~$t$ such that $r^t$ divides~$b$.
\begin{definition}
For an integer $k$, denote by $\Df(k)$ the set of prime divisors of $k$. Let~$n$ and $m$ be integers greater than $1$. For $q\in\Df(m)$, write
\[
n(q)=\begin{cases}
\displaystyle\prod_{p\in\Df(n)\setminus\{q\}} p     & \text{if $n$ is odd or $q=2$},\\[4ex]
\displaystyle 4\prod_{p\in\Df(n)\setminus\{2,q\}} p & \text{otherwise}.
\end{cases}
\]
Put
\[
b(r,n,m)=\begin{cases}
\max\limits_{q\in\Df(m)\setminus\{2\}}
\big\{\nu_2(q^2-1)+\nu_2(\ord_{n(q)}(q))-1\big\} & \text{for $r=2$},\\[2ex]
\max\limits_{q\in\Df(m)\setminus\{r\}}\big\{\nu_r(q^{r-1}-1)+\nu_r(\ord_{n(q)}(q))\big\} & \text{for $r>2$}
\end{cases}
\]
with the convention that $b(2,n,m)=2$ if $\Df(m)=\{2\}$ and $b(r,n,m)=1$ if $\Df(m)=\{r\}$ and $r>2$. We define
\[
F(n,m)=\gcd\Bigg(n,\prod_{p\in\Df(n)}p^{b(p,n,m)}\Bigg).
\]
\end{definition}
\par
Elementary number theory implies the useful fact that, if $n$ and $m$ are integers greater than $1$, then every prime divisor of $n$ is also a divisor of $F(n,m)$.
\par
The following result is the cyclic group case of \cite[Theorem~4.3]{LeuSch2005}, which generalises \cite[Theorem~5.3]{Sch1999}. We denote by $\phi(n)$ Euler's totient function.
\begin{theorem}[\cite{LeuSch2005}]
\label{thm:field_descent}
Let $G=A\times H$ be a cyclic group such that $\gcd(\abs{A},\abs{H})=1$. If $G$ contains an $(n,k,\lambda,m)$ difference set with $\gcd(m,\abs{H})=1$, then
\[
m\le \frac{\abs{H}F^2}{4\phi(F)},
\]
where $F=\gcd(\abs{A},F(n,m))$.
\end{theorem}
\par
In the case of cyclic Hadamard difference sets, we have $n=4u^2$ for $u$ odd by Theorem~\ref{thm:Hadamard_diff_set_u_odd}, so that we can always take $\abs{H}=4$ in Theorem~\ref{thm:field_descent}. It can be shown that this is always a better choice than $\abs{H}=1$. Application of Theorem~\ref{thm:field_descent} with $\abs{H}=4$ to cyclic Hadamard difference sets gives the following result (see~\cite[Corollary~4.5]{LeuSch2005}).
\begin{corollary}[\cite{LeuSch2005}]
\label{cor:field_descent_Hadamard}
If there exists a perfect binary sequence of length $4u^2$, then $u\phi(u)\le F(u^2,u)$.
\end{corollary}
\par
A combination of Corollaries~\ref{cor:self_conjugacy_bound} and~\ref{cor:field_descent_Hadamard} implies that there is no perfect binary sequence of length $4u^2$ for $1<u<11\,715$~\cite[Corollary~4.5]{LeuSch2005}. Hence we have the following result.
\begin{corollary}[\cite{LeuSch2005}]
\label{cor:nonexistence_perfect}
There is no perfect binary sequence of length $n$ for $4<n<548\,964\,900$.
\end{corollary}
\par
We illustrate the application of Corollary~\ref{cor:field_descent_Hadamard} for perfect binary sequences of length $12\,100$, which is the first case where Turyn's results~\cite{Tur1965} are insufficient to prove nonexistence.
\begin{example}
Take $n=4u^2$ for $u=55$, so that $n=12\,100$. To apply Corollary~\ref{cor:field_descent_Hadamard}, we require the value of $F(55^2,55)$. We have $\ord_{n(5)}(5)=\ord_{11}(5)=5$ and $\ord_{n(11)}(11)=\ord_{5}(11)=1$ and therefore
\begin{align*}
b(5,55^2,55)&=\nu_5(11^4-1)+\nu_5(1)=1\\
b(11,55^2,55)&=\nu_{11}(5^{10}-1)+\nu_{11}(5)=1,
\end{align*}
from which we conclude that
\[
F(55^2,55)=\gcd(55^2,5^1\cdot 11^1)=55.
\]
Since $\phi(55)=40$, by Corollary~\ref{cor:field_descent_Hadamard} the existence of a perfect binary sequence of length $12\,100$ implies $55\cdot 40\le 55$, a contradiction. Therefore there is no perfect binary sequence of length $12\,100$.
\end{example}
\par
Mossinghoff~\cite{Mos2009}, Borwein and Mossinghoff~\cite{BorMos2013}, and 
Logan and Mossinghoff~\cite{LogMos2015} proposed clever methods in order to identify numbers $n$ for which Corollary~\ref{cor:field_descent_Hadamard} does not prove nonexistence of a perfect binary sequence of length $n$. For many of these numbers, nonexistence follows from Corollary~\ref{cor:self_conjugacy_bound} or some further nonexistence results by Leung and Schmidt~\cite{LeuSch2011}, which also involve self-conjugacy arguments and the field descent method. Most notably, Leung and Schmidt~\cite{LeuSch2015} recently developed a new method, which they call the ``Anti-Field-Descent Method'', which provides further strong, albeit rather technical, nonexistence results. However, the smallest length for which the existence of a perfect binary sequences has not been decided so far is still $548\,964\,900$.
\par
We close this section with some comments on optimal binary sequences of length $n$ for $n\equiv 1\pmod 4$. Such sequences are in one-to-one correspondence with cyclic difference sets having parameters
\begin{equation}
(2u^2+2u+1,u^2,\tfrac12u(u-1),\tfrac12u(u+1))   \label{eqn:diff_set_parameters_1_mod_4}
\end{equation}
for a positive integer $u$. The cases $u=1$ and $u=2$ correspond to the binary sequences~\eqref{eqn:optimal_1_mod_4}. Turyn~\cite[p.~199]{Tur1968} reports nonexistence of these difference sets for $3\le u\le 11$. Eliahou and Kervaire~\cite{EliKer1992} used the following result due to Lander \cite[Theorem~4.5]{Lan1983} to obtain further nonexistence results.
\begin{theorem}[\cite{Lan1983}]
\label{thm:semiprimitivity_theorem}
Suppose that there exists a cyclic difference set with parameters $(n,k,\lambda,m)$. Let $d$ be a divisor of $n$ with $d>1$ and let $p$ be a prime. If $p$ is semiprimitive modulo $d$, then $p$ does not divide the square-free part of $m$. Moreover, if $d=n$, then $p$ does not divide $n$ itself.
\end{theorem}
\par
Theorem~\ref{thm:semiprimitivity_theorem} implies the nonexistence of cyclic difference sets with parameters~\eqref{eqn:diff_set_parameters_1_mod_4} for all $u$ satisfying $3\le u\le 100$, except for $u\in\{9,49,50,82\}$ (see~\cite[Table I]{EliKer1992} for details). The latter four cases can be ruled out~\cite{EliKer1992}, \cite{Bro1994} using multiplier theory. Hence there is no optimal binary sequence of length $n$ for $n\equiv 1\pmod 4$ and $13<n<20605$.
\par
Of course these results suggest a conjecture, which, to my knowledge, has not been stated explicitly in the literature.
\begin{conjecture}
There is no optimal binary sequence of length $n>13$ for $n\equiv 1\pmod 4$. Equivalently there is no cyclic difference set with parameters~\eqref{eqn:diff_set_parameters_1_mod_4} for~$u>2$.
\end{conjecture}


\section{Aperiodic autocorrelation of binary sequences}
\label{sec:aperiodic}

\subsection{Barker sequences}
\label{sec:barker}

For every binary sequence $A$ of length $n$, the aperiodic autocorrelation $C_u(A)$ is an integer with  parity $n-u$. A \emph{Barker sequence} is a binary sequence with the ideal property that all nontrivial aperiodic autocorrelations are either $0$ or $1$ in magnitude. (Barker's original definition~\cite{Bar1953} requires that all nontrivial aperiodic autocorrelations are either $0$ or $-1$, but it has become customary to impose our slightly less restrictive condition.)
\par
Notice that for fixed $a,b\in\{0,1\}$, the transformation $A(k)\mapsto A(k)(-1)^{a+bk}$ preserves the Barker property. We can therefore assume  without loss of generality that a Barker sequence $A$ satisfies $A(1)=A(2)=1$. The only known Barker sequences with this property are (writing $+$ for $1$ and~$-$~for~$-1$)
\begin{alignat*}{3}
n&=2&:  & \qquad (+\,+),\\[.5ex]
n&=3&:  & \qquad (++-),\\[.5ex]
n&=4&:  & \qquad (+++\,-),\qquad (++-\,+),\\[.5ex]
n&=5&:  & \qquad (+++-+),\\[.5ex]
n&=7&:  & \qquad (+++--+-),\\[.5ex]
n&=11&: & \qquad (+++---+--+-),\\[.5ex]
n&=13&: & \qquad (+++++--++-+-+).
\end{alignat*}
Indeed, it has been conjectured since at least 1960~\cite{Tur1960} that there are no other lengths for which Barker sequences exist.
\begin{conjecture}[\cite{Tur1960}]
\label{con:Barker}
There is no Barker sequence of length greater than~$13$.
\end{conjecture}
\par
This conjecture is known to be true for sequences of odd length, as proven by Turyn and Storer~\cite{TurSto1961}. A simpler proof was recently given by Schmidt and Willms~\cite{SchWil2015}.
\begin{theorem}[\cite{TurSto1961},~\cite{SchWil2015}]
\label{thm:Barker_odd_length}
There is no Barker sequence of odd length greater than~$13$.
\end{theorem}
\par
Indeed, the case that the length is odd in Conjecture~\ref{con:Barker} appears to be considerably easier than the case of even length for the following reason. Since exactly one of $u$ or $n-u$ is odd for odd $n$, it follows from~\eqref{eqn:R_from_C} and Lemma~\ref{lem:Ru_mod_four} that a Barker sequence~$A$ of odd length $n$ satisfies 
\[
C_u(A)=\begin{cases}
0              & \text{for even $u>0$}\\
(-1)^{(n-1)/2} & \text{for odd $u$}.
\end{cases}
\]
This fixes all aperiodic autocorrelations of a Barker sequence of odd length, which is the key to prove Theorem~\ref{thm:Barker_odd_length}. A similar reasoning implies that a Barker sequence of even length greater than $2$ must have length a multiple of $4$ and all of its nontrivial periodic autocorrelations equal to zero. Hence we have the following.
\begin{proposition}
\label{pro:Barker_perfect}
Every Barker sequence of even length greater than $2$ is a perfect binary sequence.
\end{proposition}
\par
In view of Proposition~\ref{pro:Barker_perfect} and Theorem~\ref{thm:Barker_odd_length}, Turyn's Conjecture~\ref{con:Barker} is implied by Ryser's Conjecture~\ref{con:perfect_binary_sequences} and all nonexistence results for perfect binary sequences immediately provide nonexistence results for Barker sequences. In particular, by Theorem~\ref{thm:Hadamard_diff_set_u_odd}, the length of a Barker sequence of even length greater than $2$ is four times an odd square and, by Corollary~\ref{cor:nonexistence_perfect}, there is no Barker sequence of even length $n$ for $4<n<548\,964\,900$. The only known nonexistence result for Barker sequences of even length that is not implied by results for perfect binary sequences is the following result due to Eliahou, Kervaire, and Saffari~\cite{EliKerSaf1991} (which as explained in~\cite{EliKerSaf1991} follows from the forthcoming Proposition~\ref{pro:Golay_pmod4}). 
\begin{theorem}[\cite{EliKerSaf1991}]
\label{thm:Barker_3_mod_4}
If a Barker sequence of even length $n$ exists, then every odd prime divisor of $n$ is congruent to $3$ modulo $4$.
\end{theorem}
\par
As shown by Mossinghoff~\cite{Mos2009}, the combination of Corollaries~\ref{cor:self_conjugacy_bound} and~\ref{cor:field_descent_Hadamard} and Theorem~\ref{thm:Barker_3_mod_4} implies that there is no Barker sequence of even length $n$ for 
\[
4<n<189\,260\,468\,001\,034\,441\,522\,766\,781\,604.
\]
Refined methods by Borwein and Mossinghoff~\cite{BorMos2013} and Leung and Schmidt~\cite{LeuSch2011},~\cite{LeuSch2015} imply the following slightly stronger result.
\begin{proposition}
There is no Barker sequence of even length $n$ for $4<n\le 4\cdot 10^{33}$.
\end{proposition}
\par
As noted by Leung and Schmidt~\cite{LeuSch2015}, there are currently $8125$ known numbers $n$ up to $10^{100}$ for which the known methods fail to settle the nonexistence of a Barker sequence of length $n$. The smallest of these numbers is larger than $10^{51}$, namely $4u^2$ for $u=30109\cdot 1128713\cdot 2167849\cdot 268813277$. However it is not clear whether these known open cases are exhaustive for $n\le 10^{100}$.


\subsection{Measures of smallness of aperiodic autocorrelations}
\label{sec:measures}

In response to the presumed nonexistence of long Barker sequences, several authors have studied different measures for the collective smallness of the aperiodic autocorrelations of sequences. For a sequence $A$ of length $n$ and a real number $r>0$, define
\begin{gather*}
M_r(A)=\Bigg(\sum_{0<u<n}\,\abs{C_u(A)}^r\Bigg)^{1/r}
\intertext{and}
M(A)=\max_{0<u<n}\,\abs{C_u(A)},
\end{gather*}
which equals the limit of $M_r(A)$ as $r\to\infty$. We are interested in minimising these functions over the set of binary sequences of a given length. Accordingly, we define the arithmetic functions
\begin{equation}
m_r(n)=\min_{A\in\mathfrak{B}_n}M_r(A)   \label{eqn:def_mr}
\end{equation}
and
\begin{equation}
m(n)=\min_{A\in\mathfrak{B}_n}M(A),   \label{eqn:m-function}
\end{equation}
where $\Bf_n$ is the set of binary sequences of length $n$. The principal problem is to understand the behaviour of these functions as $n$ tends to infinity.
\par
Two measures have received particular attention: $M(A)$, called the \emph{peak sidelobe level} of $A$, and $M_2(A)$, which is essentially the sum of squares of the nontrivial aperiodic autocorrelations of $A$. 
\par
In Section~\ref{sec:random_binary_sequences}, we shall see how probabilistic methods help to understand the asymptotic behaviour of the functions $m(n)$ and $m_r(n)$. In Sections~\ref{sec:PSL} and~\ref{sec:merit_factor}, we study the measures $M(A)$ and $M_2(A)$, respectively, where our focus is in particular on constructive results.


\subsection{Random binary sequences}
\label{sec:random_binary_sequences}

In this section our goal is to obtain information on the growth rate of the functions $m(n)$ and $m_r(n)$ using probabilistic methods. As before, $\mathfrak{B}_n$ denotes the set of binary sequences of length $n$ and, throughout this section, $A_n$ is drawn at random from $\mathfrak{B}_n$, equipped with the uniform probability measure. In other words, each of the $n$ entries in $A_n$ is drawn independently from $\{-1,1\}$ with $\Pr(-1)=\Pr(1)=1/2$. By $\E(X)$ we denote the expectation of a random variable $X$.
\par
We are interested in the asymptotic behaviour, as $n\to\infty$, of $M(A)$ and $M_r(A)$ for most binary sequences $A$ of length $n$. This problem was first studied by Moon and Moser~\cite{MooMos1968} for the peak sidelobe level $M(A)$. In particular, they asked for arithmetic functions $L(n)$ and $U(n)$ such that
\[
\lim_{n\to\infty}\Pr\Big[L(n)\le M(A_n)\le U(n)\Big]=1.
\]
This implies in particular that $m(n)$ grows not faster than $U(n)$. 
\par
Some nontrivial results for such functions $L(n)$ and $U(n)$ were given by Moon and Moser~\cite{MooMos1968}, which were later improved by Mercer~\cite{Mer2006} and Alon, Litsyn, and Shpunt~\cite{AloLitShp2010}. Further improvements by the author~\cite{Sch2014} show that we can in fact take 
\[
L(n)=(1-\e)\sqrt{2n\log n}\quad \text{and} \quad U(n)=(1+\e)\sqrt{2n\log n}
\]
for an arbitrary $\e>0$. To state the result slightly more formally, recall that a sequence of random variables $X_1,X_2,\dots$ \emph{converges in probability} to a constant $c$ if 
\[
\Pr[\abs{X_n-c}>\e]\to 0
\]
as $n\to\infty$ for all $\e>0$.
\begin{theorem}[\cite{Sch2014}]
\label{thm:psl_dist}
Let $A_n$ be drawn at random from $\mathfrak{B}_n$, equipped with the uniform probability measure. Then, as $n\to\infty$,
\[
\frac{M(A_n)}{\sqrt{n\log n}}\to\sqrt{2}\quad\mbox{in probability}
\]
and
\[
\frac{\E(M(A_n))}{\sqrt{n\log n}}\to\sqrt{2}.
\]
\end{theorem}
\par
In~\cite{Sch2012a}, the following complementary result for $M_r(A_n)$ was proved, in which $\Gamma(z)=\int_0^\infty e^{-t}t^{z-1}\,dt$ denotes the \emph{gamma function}, satisfying $\Gamma(p+1)=p!$ when $p$ is a nonnegative integer.
\begin{theorem}{\cite{Sch2012a}}
\label{thm:norms_dist}
Let $A_n$ be drawn at random from $\mathfrak{B}_n$, equipped with the uniform probability measure, and let $r$ be a positive real number. Then, as $n\to\infty$,
\[
\frac{M_r(A_n)}{n^{1/2+1/r}}
\to \bigg(\frac{\Gamma(r+1)}{2^{r/2}\,\Gamma(r/2+2)}\bigg)^{1/r}\quad\text{in probability}
\]
and
\begin{equation}
\frac{\E(M_r(A_n)^r)}{n^{r/2+1}}\to\frac{\Gamma(r+1)}{2^{r/2}\,\Gamma(r/2+2)}.    \label{eqn:lim_Er}
\end{equation}
Moreover, for $r\ge 1$, as $n\to\infty$,
\[
\frac{\E(M_r(A_n))}{n^{1/2+1/r}}\to \bigg(\frac{\Gamma(r+1)}{2^{r/2}\,\Gamma(r/2+2)}\bigg)^{1/r}.
\]
\end{theorem}
\par
Theorems~\ref{thm:psl_dist} and~\ref{thm:norms_dist} provide upper bounds for the growth rate of the functions $m(n)$ and $m_r(n)$, namely
\begin{equation}
\limsup_{n\to\infty}\frac{m(n)}{\sqrt{n\log n}}\le \sqrt{2}   \label{eqn:growth_m}
\end{equation}
and
\begin{equation}
\limsup_{n\to\infty}\frac{m_r(n)}{n^{1/2+1/r}}\le \bigg(\frac{\Gamma(r+1)}{2^{r/2}\,\Gamma(r/2+2)}\bigg)^{1/r}.   \label{eqn:growth_mr}
\end{equation}
In Section~\ref{sec:PSL}, we provide an explicit construction, which shows that~\eqref{eqn:growth_m} can be improved to
\[
m(n)\le \sqrt{2n\log (2n)}\quad\text{for all $n>1$}.
\]
For finite $r\ne 2$, nothing stronger than~\eqref{eqn:growth_mr} is known, and for $r=2$, the best known result is
\[
\limsup_{n\to\infty}m_2(n)/n\le c,
\]
where $c<25/89$ is strictly smaller than $1/\sqrt{2}$ (see the forthcoming Corollary~\ref{cor:Legendre}).
\par
When $r$ is a positive integer, the exact values of $\E(M_r(A_n)^r)$ are known. Since a random variable cannot always exceed its expected value, such values give bounds for $m_r(n)$ for integral $r$ and specific values of~$n$. Mercer~\cite{Mer2006} showed that, when $r$ is an even positive integer, then $\E(M_r(A_n)^r)$ is a polynomial of degree $r/2+1$ in $n$. By~\eqref{eqn:lim_Er}, the leading coefficient of this polynomial is 
\[
\frac{r!}{2^{r/2}\,(r/2+1)!}.
\]
For example,
\begin{align}
\E(M_2(A_n)^2)&=\tfrac{1}{2}(n^2-n),  \label{eqn:mean_m2}\\
\E(M_4(A_n)^4)&=\tfrac{1}{2}(2n^3-5n^2+3n).   \nonumber
\end{align}
The author showed~\cite{Sch2012a} that, when $r$ is an odd positive integer, then
\[
\frac{4^n}{{2n\choose n}}\E(M_r(A_{2n})^r)\quad\text{and}\quad \frac{4^n}{{2n\choose n}}\E(M_r(A_{2n+1})^r)
\]
are polynomials of degree $(r+3)/2$ in $n$. It can be deduced from~\eqref{eqn:lim_Er} that the leading term of these polynomials is
\[
\frac{2^{r+2}\,(\frac{r-1}{2})!}{r+2}.
\]
For example,
\begin{align*}
\E(M_1(A_{2n}))    &=\binom{2n}{n}\frac{8n^2-2n}{3\cdot 4^n},\\
\E(M_1(A_{2n+1}))  &=\binom{2n}{n}\frac{8n^2+4n}{3\cdot 4^n},\\
\E(M_3(A_{2n})^3)  &=\binom{2n}{n}\frac{96n^3-68n^2+2n}{15\cdot 4^n},\\
\E(M_3(A_{2n+1})^3)&=\binom{2n}{n}\frac{96n^3+52n^2+2n}{15\cdot 4^n}.
\end{align*}


\subsection{The peak sidelobe level of binary sequences}
\label{sec:PSL}

In this section we continue to study the function $m(n)$, defined in~\eqref{eqn:m-function}. We are in particular interested in constructive existence results. The value of $m(n)$ has been determined via exhaustive search for all $n\le 80$ (see~\cite{LeuPot2014} for the latest results). Many authors have put considerable computational effort in finding binary sequences with small peak sidelobe level (see Nunn and Coxson~\cite{NunCox2008}, for example), showing that the function $m(n)$ satisfies
\begin{alignat}{2}
m(n)&\le 1 && \quad\text{for each $n\le 5$},   \nonumber\\
m(n)&\le 2 && \quad\text{for each $n\le 21$},   \nonumber\\
m(n)&\le 3 && \quad\text{for each $n\le 48$},   \label{eqn:values_of_m}\\
m(n)&\le 4 && \quad\text{for each $n\le 82$},   \nonumber\\
m(n)&\le 5 && \quad\text{for each $n\le 105$}.   \nonumber
\end{alignat}
Turyn conjectured~\cite{Tur1963},~\cite[p.~198]{Tur1968} that the infimum limit of $m(n)$ is infinite. Ein-Dor, Kanter, and Kinzel~\cite{EinKanKin2002} used a heuristic argument to obtain an ``educated guess'' about the growth of the function $m(n)$. We summarise their results in the following form.
\begin{conjecture}
\label{con:min_PSL}
As $n\to\infty$, we have
\[
\frac{m(n)}{\sqrt{n}}\to d,\quad\text{where}\quad d=0.435\dots\,.
\] 
\end{conjecture}
\par
If $A_n$ is drawn from the set of binary sequences of length $n$, equipped with the uniform probability measure, then there are dependencies among the random variables
\begin{equation}
\frac{C_1(A_n)}{\sqrt{n-1}},\,\frac{C_2(A_n)}{\sqrt{n-2}},\,\dots,\,\frac{C_{n-1}(A_n)}{\sqrt{1}}.   \label{eqn:normalised_correlations}
\end{equation}
The underlying heuristic assumption leading to the conclusion of Conjecture~\ref{con:min_PSL} is to treat~\eqref{eqn:normalised_correlations} as mutually independent standard normal random variables. The normality is partly justified by the central limit theorem. The independence is also partly justified: The methods used to prove~\cite[Proposition~7]{Sch2012a} can be used to show that, for a \emph{fixed} positive integer $v$, the random vector
\[
\bigg(\frac{C_1(A_n)}{\sqrt{n-1}},\,\frac{C_2(A_n)}{\sqrt{n-2}},\,\dots,\,\frac{C_v(A_n)}{\sqrt{n-v}}\bigg)
\]
converges in distribution to a multivariate normal distribution with identity covariance matrix. The known values of $m(n)$ also lend evidence in favour of Conjecture~\ref{con:min_PSL}. Writing 
$f(x)=0.435\sqrt{x}$, then $f(x)-1$ changes sign for $x\in(5,6)$, $f(x)-2$ changes sign for $x\in(21,22)$, and $f(x)-3$ changes sign for $x\in(47,48)$. This should be compared with the data in~\eqref{eqn:values_of_m}.
\par
In the remainder of this section, we discuss constructive results. In~\cite{Sch2012} the author gives a construction for a binary sequence of length~$n$ with peak sidelobe level at most $\sqrt{2n\log(2n)}$ for every $n>1$, thus showing that
\[
m(n)\le \sqrt{2n\log(2n)}.
\]
The construction is inspired by a method in probabilistic combinatorics, known as derandomisation.
\begin{construction}[\cite{Sch2012}]
\label{constr:sum_of_cosh}
\sloppypar
Let $n$ be a positive integer and construct a binary sequence $B_n$ of length $n$ recursively by
\[
B_n(k)=-\sign\Bigg[\sum_{u=1}^{k-1}B_n(k-u)\sinh\bigg(\sqrt{\frac{2\log(2n)}{n}}\;\;\sum_{j=0}^{k-u-1}B_n(j)B_n(j+u)\bigg)\Bigg],
\]
where, by convention, $\sign(0)=-1$.
\end{construction}
\par
As shown in~\cite{Sch2012}, the sequence $B_n$ can be efficiently constructed with $O(n^2)$ multiplications and additions. 
\begin{theorem}[\cite{Sch2012}]
\label{thm:sum_of_cosh}
The binary sequence $B_n$ of length $n>1$ obtained under Construction~\ref{constr:sum_of_cosh} satisfies $M(B_n)\le \sqrt{2n\log (2n)}$.
\end{theorem}
\par
Theorem~\ref{thm:sum_of_cosh} gives the currently best known upper bound for infinitely many values of $m(n)$, although it guarantees only a peak sidelobe level of roughly the same as that of a typical binary sequence (see Theorem~\ref{thm:psl_dist}). Numerical results~\cite{Sch2012} however lend evidence to the following conjecture.
\begin{conjecture}[{\cite{Sch2012}}]
\label{con:growth}
Let $B_n$ be the binary sequence of length $n$ obtained under Construction~\ref{constr:sum_of_cosh}. Then there exist positive constants $c_1$ and $c_2$ such that, for all~$n>1$,
\[
c_1\sqrt{n\log\log n}\le M(B_n)\le c_2\sqrt{n\log\log n}.
\]
\end{conjecture}
\par
Some examples for small $n$ reveal that, if $c_2$ in Conjecture~\ref{con:growth} exists, then $c_2$ must be strictly greater than $1$. It is however conceivable that
\[
\limsup_{n\to\infty}\frac{M(B_n)}{\sqrt{n\log\log n}}\le 1.
\]
The correctness of Conjecture~\ref{con:growth} implies that the sequences $B_n$ are exceptional in the sense that their peak sidelobe level grows strictly more slowly than that of most binary sequences, as given in Theorem~\ref{thm:psl_dist}. 
\par
Further candidates of families of binary sequences whose peak sidelobe grows more slowly than that of most binary sequences are Legendre sequences and Galois sequences (see Section~\ref{sec:bounds_and_constructions}), although the currently known proven results are not as strong as those in Theorem~\ref{thm:sum_of_cosh}. 
\par
A \emph{cyclic shift} by $r$ elements of a sequence $A$ of length~$n$ is the sequence of length~$n$ whose $k$-th entry is $A(k+r)$, where as usual the index is reduced modulo $n$. Note that, while the periodic autocorrelations remain unchanged for all cyclic shifts of a sequence, the aperiodic autocorrelations can vary considerably over the cyclic shifts of a sequence.
\par
For Legendre sequences, the following result was proved by Mauduit and S{\'a}rk{\"o}zy~\cite{MauSar1997}.
\begin{theorem}[\cite{MauSar1997}]
\label{thm:psl_legendre}
The peak sidelobe level of every cyclic shift of a Legendre sequence of (prime) length $p$ is at most $1+18\sqrt{p}\log p$.
\end{theorem}
\par
Numerical investigations by Boehmer~\cite{Boe1967}, Turyn~\cite[p.~203]{Tur1968}, and in particular by Jedwab and Yoshida~\cite{JedYos2006} suggest that the bound of Theorem~\ref{thm:psl_legendre} can be improved, perhaps to a small constant times $\sqrt{p\log p}$.
\par
For Galois sequences, the following result was proved by Sarwate~\cite{Sar1984}.
\begin{theorem}[{\cite{Sar1984}}]
\label{thm:psl_galois}
The peak sidelobe level of a Galois sequence of length $n=2^m-1$ is at most $1+(2/\pi)\sqrt{n+1}\log(4n/\pi)$.
\end{theorem}
\par
Note that every cyclic shift of a Galois sequence is also a Galois sequence, so Theorem~\ref{thm:psl_galois} also applies to all cyclic shifts of a Galois sequence.
\par
We shall see in Theorem~\ref{thm:Galois} that the asymptotic behaviour of $M_2(A)$ is known for Galois sequences. A combination with the standard norm inequality 
\[
M(A)\,n^{1/2}\ge M_2(A),
\]
valid for arbitrary sequences $A$ of length $n>1$, implies the following asymptotic lower bound.
\begin{theorem}
\label{thm:PSL_Galois_lower_bound}
Let $n$ take values only in the set of Mersenne numbers and let $Y_n$ be a Galois sequence of length~$n$. Then
\[
\liminf_{n\to\infty}\frac{M(Y_n)}{n^{1/2}}\ge \frac{1}{\sqrt{6}}.
\]
\end{theorem}
\par
A similar result can also be established for Legendre sequences. Theorems~\ref{thm:psl_galois} and~\ref{thm:PSL_Galois_lower_bound} determine the asymptotic behaviour of the peak sidelobe level of Galois sequences up to a factor of roughly $\log n$. Numerical results suggest that the upper bound in Theorem~\ref{thm:psl_galois} can be improved. In particular, extensive numerical investigations by Dmitriev and Jedwab~\cite{DmiJed2007} give evidence supporting the following conjecture.
\begin{conjecture}
\label{con:psl_galois}
The peak sidelobe level of a Galois sequence of length $n=2^m-1$ is at most $C\sqrt{n}\,\log\log n$ for some absolute constant $C$.
\end{conjecture}
\par
The correctness of Conjecture~\ref{con:psl_galois} implies that Galois sequences are exceptional in the sense that their peak sidelobe level grows strictly more slowly than that of most binary sequences. 
\par
Even more striking observations can be obtained from a numerical analysis of random Galois sequences. It is well known that there are exactly $n\phi(n)/m$ Galois sequences of length $n=2^m-1$, where $\phi(n)$ is Euler's totient function. Numerical investigations by Dmitriev and Jedwab~\cite{DmiJed2007} lend strong evidence to the following conjecture.
\begin{conjecture}
\label{con:psl_random_galois}
Let $n$ take values only in the set of Mersenne numbers. Let $Y_n$ be drawn from the set of Galois sequences of length $n$, equipped with the uniform probability measure, and let $W(Y_n)$ be the maximum peak sidelobe level over all cyclic shifts of $Y_n$. Then the limit
\[
\lim_{n\to\infty}\frac{\E(W(Y_n))}{\sqrt{n}}
\]
exists and is finite.
\end{conjecture}
\par
Indeed it has been verified in~\cite{DmiJed2007} that, with the notation as in Conjecture~\ref{con:psl_random_galois}, the value $\E(W(Y_n))/\sqrt{n}$ lies within $3\%$ of $1.31$ for all values of~$m$ between $13$ and $25$. The correctness of Conjectures~\ref{con:min_PSL} and~\ref{con:psl_random_galois} would imply that the family of Galois sequences contains a subfamily whose peak sidelobe level is nearly optimal. 


\subsection{The merit factor of binary sequences}
\label{sec:merit_factor}

In this section we are interested in the measure $M_2(A)$ for binary sequences $A$, or equivalently, in the sum of squares of the nontrivial aperiodic autocorrelations of binary sequences. For a sequence $A$ of length~$n$, it is customary to study the normalised measure
\[
F(A)=\frac{C_0(A)^2}{\displaystyle 2\sum_{0<u<n}\abs{C_u(A)}^2}
\]
(provided that the denominator is nonzero), which Golay~\cite{Gol1972} called the \emph{merit factor} of~$A$. A large merit factor means that the sum of squares of the nontrivial autocorrelations is small when compared to the squared trivial autocorrelation (which always equals $n^2$ for binary sequences of length $n$).
\par
The determination of the largest possible merit factor of long binary sequences is of considerable importance in various contexts. In digital communications, binary sequences with large merit factor correspond to signals whose energy is very uniformly distributed over frequency~\cite{BeeClaHer1985}. In theoretical physics, binary sequences achieving the largest merit factor for their length correspond to the ground states of Bernasconi's Ising spin model~\cite{Ber1987}. The growth rate of the optimal merit factor of binary sequences, as the sequence length increases, is related to classical conjectures due to Littlewood~\cite{Lit1966},~\cite{Lit1968} and Erd\H{o}s~\cite{Erd1995},~\cite{NewByr1990} on the asymptotic behaviour of norms of polynomials on the unit circle. 
\par
This latter relationship arises because, when a sequence~$A$ of length $n$ is represented as a polynomial $f_A(z)=\sum_{k=0}^{n-1}A(k)z^k$, its merit factor $F(A)$ satisfies
\[
F(A)=\frac{\norm{f_A}_2^4}{\norm{f_A}_4^4-\norm{f_A}_2^4},
\]
where, for $1\le\alpha<\infty$,
\[
\norm{f_A}_\alpha=\left(\frac{1}{2\pi} \int_0^{2 \pi} \bigabs{f_A(e^{i\phi})}^\alpha \,d\phi\right)^{1/\alpha}
\]
is the $L^\alpha$ norm on the unit circle of the polynomial $f_A(z)$. Note that $\norm{f_A}_2=\sqrt{n}$ if $A$ is a unimodular sequence of length $n$. There is an extensive body of research dealing with extremal problems for such norms (see~\cite{Bor2002} for a survey of selected problems).
\par
Define
\[
\varphi(n)=\max_{A\in\mathfrak{B}_n}F(A),
\]
where $\mathfrak{B}_n$ is the set of binary sequences of length $n$. This function is related to the function $m_2(n)$, defined in~\eqref{eqn:def_mr}, via $2\varphi(n)=(n/m_2(n))^2$. It follows from~\eqref{eqn:mean_m2} that, when $A$ is drawn uniformly at random from $\mathfrak{B}_n$, then $\E(1/F(A))=1-1/n$, which gives a lower bound for $\varphi(n)$. Various conjectures on the asymptotic behaviour of $\varphi(n)$ have appeared in the literature. We mention in particular two contradicting conjectures by Golay~\cite{Gol1982} and Littlewood~\cite{Lit1966}.
\begin{conjecture}[\cite{Gol1982}]
\label{con:mf_golay}
$\lim_{n\to\infty}\varphi(n)$ exists and equals $12.32\dots$.
\end{conjecture}
\par
\begin{conjecture}[\cite{Lit1966}]
\label{con:mf_littlewood}
$\limsup_{n\to\infty}\varphi(n)=\infty$.
\end{conjecture}
\par
Conjecture~\ref{con:mf_golay} uses the same heuristic reasoning as that leading to Conjecture~\ref{con:min_PSL} for the minimum peak sidelobe level. Apparently, Conjecture~\ref{con:mf_littlewood} is based solely on (very limited) numerical data.
\par
In view of the above conjectures, it is interesting that Fredman, Saffari, and Smith~\cite{FreSafSmi1989} proved that symmetric binary sequences have bounded merit factor. In particular,~\cite{FreSafSmi1989} contains the following more precise result.
\begin{theorem}[\cite{FreSafSmi1989}]
Let $A$ be a unimodular sequence of length $n$ satisfying $A(n-1-k)=\overline{A(k)}$ for $0\le k<n$. Then
\[
\frac{1}{F(A)}\ge \sup_{\lambda>0}\:\frac{1}{\cosh{(2\lambda)}}\bigg(\frac{(\sinh\lambda)^2}{\lambda^2}-1\bigg).
\]
In particular, $F(A)<9.55$.
\end{theorem}
\par
The strongest existence result that Littlewood was able to prove~\cite{Lit1968} arises from a construction due to Shapiro~\cite{Sha1951},~\cite{Rud1959}. Take $A_0=B_0=(1)$ and construct binary sequences $A_m$ and $B_m$ of length $2^m$ recursively with the rule
\begin{equation}
A_{m+1}=(A_m,\,B_m)\quad\text{and}\quad B_{m+1}=(A_m,\,-B_m).   \label{eqn:shapiro_recursion}
\end{equation}
The sequences $A_m$ and $B_m$ are called the \emph{Shapiro sequences} of length $2^m$.
\begin{theorem}[\cite{Lit1968}]
\label{thm:mf_shapiro}
Let $A_m$ be either Shapiro sequence of length $2^m$.~Then
\[
F(A_m)=\frac{3}{1-(-1/2)^m}.
\]
In particular, the asymptotic merit factor as $m\to\infty$ of Shapiro sequences equals $3$.
\end{theorem}
\par
Theorem~\ref{thm:mf_shapiro} was first proved by Littlewood~\cite[Chapter~III, Problem~19]{Lit1968}, but was later obtained independently by H{\o}holdt, Jensen, and Justesen~\cite{HohJenJus1985} and Newman and Byrnes~\cite{NewByr1990}. Merit factors of generalisations of Shapiro sequences have also been studied in~\cite{HohJenJus1985} and~\cite{BorMos2000}.
\par
The result of Theorem~\ref{thm:mf_shapiro} was subsequently improved by H{\o}holdt and Jensen~\cite{HohJen1988} and by Jedwab, Katz, and Schmidt~\cite{JedKatSch2013a} (see also~\cite{JedKatSch2013}) using Legendre sequences. In order to state the results, we require the following notation. Let $A$ be a sequence of length~$n$. Let $r$ and $t$ be integers that can depend on $n$, where $t>0$, and define the sequence $A^{r,t}$ to be the sequence of length $t$ whose $k$-th entry is $A(k+r)$, where as usual the index in $A(k+r)$ is reduced modulo $n$. Informally, the sequence $A^{r,t}$ is obtained from $A$ by cyclically permuting (shifting) the sequence elements through $r$ positions, and then truncating when $t<n$ or periodically extending (appending) when $t>n$. For example, if $A=(1,1,-1)$, then $A^{1,4}=(1,-1,1,1)$.
\par
Define the function $g:\R\times \R^+\to\R$ by
\[
\frac{1}{g(R,T)}=1-\frac{4T}{3}+4\sum_{m\in\N}\max\bigg(0,1-\frac{m}{T}\bigg)^2+\sum_{m\in\Z}\max\bigg(0,1-\biggabs{1+\frac{2R-m}{T}}\bigg)^2,
\]
where $\N$ is the set of positive integers.
\begin{theorem}[{\cite{JedKatSch2013a}}]
\label{thm:Legendre}
Let $X_p$ be the Legendre sequence of length $p$ and let $R$ and $T>0$ be real. If $r/p\to R$ and $t/p\to T$ as $p\to\infty$, then $F(X_p^{r,t})\to g(R,T)$ as $p\to\infty$.
\end{theorem}
\par
The case $T=1$ of Theorem~\ref{thm:Legendre} implies that $X_p^{r,p}$ has asymptotic merit factor $g(R,1)$ if $r/p \to R$ as $p\to\infty$. Since
\[
\frac{1}{g(R,1)}=\frac{1}{6}+8\bigg(\abs{R}-\frac{1}{4}\bigg)^2\quad\text{for $\abs{R}\le \frac{1}{2}$},
\]
the maximum asymptotic merit factor that can be attained in this way is~$g(1/4,1)=6$. This recovers the result by H{\o}holdt and Jensen~\cite{HohJen1988}, which was mentioned above. 
\par
The function $g$ satisfies $g(R,T)=g(R+1/2,T)$ on its entire domain. As shown in~\cite[Corollary 3.2]{JedKatSch2013a}, the global maximum of $g(R,T)$ exists and equals
\begin{equation}
\text{$6.342061\dots$, the largest root of $29x^3-249x^2+417x-27$}.   \label{eqn:Fa}
\end{equation}
The global maximum is unique for $R\in[0,1/2)$, and is attained when $T=1.057827\dots$ is the middle root of $4x^3-30x+27$ and $R=3/4-T/2$. We therefore obtain the following consequence of Theorem~\ref{thm:Legendre}.
\begin{corollary}[{\cite{JedKatSch2013a}}]
\label{cor:Legendre}
There exist binary sequences $B_1,B_2,\dots$ of strictly increasing length satisfying $F(B_n)\to \Fc$ as $n\to\infty$, where $\Fc$ is given in~\eqref{eqn:Fa}. 
\end{corollary}
\par
Corollary~\ref{cor:Legendre} gives the currently best known result on the asymptotic merit factor of binary sequences. However, some numerical experiments by Baden~\cite{Bad2011} suggest strongly that~\eqref{eqn:Fa} is not the value of $\limsup_{n\to\infty}\varphi(n)$. 
\par
Theorem~\ref{thm:Legendre} has been generalised in various ways. First,~\cite[Theorem~2.1]{JedKatSch2013} establishes that certain binary sequences of length $2p$ and $4p$ constructed from a Legendre sequence of length $p$ have essentially the same asymptotic merit factor as Legendre sequences. Second,~\cite[Theorem~2.3]{JedKatSch2013} generalises Theorem~\ref{thm:Legendre} in the sense that one can include also binary sequences of composite lengths whose entries are derived from the Jacobi symbol. The third generalisation is \cite[Theorem~2.3]{GunSch2015} and more far-reaching. This result considers the characteristic sequences of subsets of $\F_p$ obtained by joining $m/2$ of the $m$ cyclotomic classes in $\F_p^*$ of (even) order $m$, where $p$ satisfies $p\equiv 1\pmod m$. For $m=2$, we can obtain Legendre sequences, but several other popular sequence families also arise in this way~\cite{Hal1956},~\cite{AraDinHelKumMar2001},~\cite{DinHelLam1999}, including the characteristic sequences of Hall difference 
sets. The asymptotic merit factor of these sequences is determined in \cite{GunSch2015} subject to a condition involving the asymptotic behaviour of their periodic autocorrelations. This condition can be checked using cyclotomic numbers and usually imposes restrictions on the underlying prime numbers.
\par
The asymptotic behaviour of the merit factor of sequences related to Galois and Sidelnikov sequences is also known. To state the results, we require the function $h:\R^+\to\R$ given by
\[
\frac{1}{h(T)}=1-\frac{2T}{3}+4\sum_{m\in\N}\max\bigg(0,1-\frac{m}{T}\bigg)^2.
\]
\begin{theorem}[{\cite{JedKatSch2013},~\cite{GunSch2015}}]
\label{thm:Galois}
Let $q$ take values only in the set of prime powers. Let $Y_q$ be either a Sidelnikov or a Galois sequence of length $q-1$ (depending on whether $q$ is odd or even). Let $T>0$ be real. If $t/q\to T$ as $q\to\infty$, then $F(Y_q^{r,t})\to h(T)$ as $q\to\infty$.
\end{theorem}
\par
The case $T=1$ of Theorem~\ref{thm:Galois} implies that $Y_q^{r,q-1}$ has asymptotic merit factor $h(1)=3$, which was already proved by Jensen, Jensen, and H{\o}holdt~\cite{JenJenHoh1991} for Galois sequences. The general result was first obtained by Jedwab, Katz, and Schmidt~\cite{JedKatSch2013}  for Galois sequences and by G\"unther and Schmidt~\cite{GunSch2015} for Sidelnikov sequences. It was also shown in~\cite{GunSch2015} that the conclusion of Theorem~\ref{thm:Galois} remains true if Galois sequences are replaced by the characteristic sequences of cyclic difference sets obtained using the Gordon-Mills-Welch construction (see Section~\ref{sec:difference_sets}).
\par
As shown in~\cite{JedKatSch2013}, the global maximum of $h(T)$ exists and equals
\[
\text{$3.342065\dots$, the largest root of $7x^3-33x^2+33x-3$}.
\]
The global maximum is unique and is attained for $T=1.115749\dots$, which is the middle root of $x^3-12x+12$. 


\section{Autocorrelation of nonbinary sequences}
\label{sec:nonbinary}

In this section we study properties of periodic and aperiodic autocorrelations of \emph{$H$-phase} sequences, which are sequences whose entries are $H$-th roots of unity, and more generally of \emph{unimodular} sequences, which are sequences whose entries have unit magnitude.

\subsection{Periodic autocorrelation of nonbinary sequences}

We have seen in Section~\ref{sec:Nonexistence_perfect} that it seems unlikely that there is a perfect binary sequence of length greater than $4$. This prompts the question as to whether there are perfect  $H$-phase sequences, namely $H$-phase sequences whose nontrivial periodic autocorrelations are all zero, for larger lengths and some $H>2$. We shall see that perfect $H$-phase sequences do exist for all lengths if we allow $H$ to grow with the length.
\par
Indeed, Mow~\cite{Mow1995} proposed the following generalisation of Conjecture~\ref{con:perfect_binary_sequences}.
\begin{conjecture}[\cite{Mow1995}]
\label{con:perfect_sequences}
Let $n$ be an integer with $n>1$ and square-free part~$r$. Then a perfect $H$-phase sequence of length $n$ exists if and only if $H$ is divisible by
\[
\begin{cases}
2\sqrt{rn} & \text{for $n\equiv 2\pmod 4$}\\[.5ex]
\sqrt{rn}  & \text{otherwise}.
\end{cases}
\]
\end{conjecture}
\par
A perfect $n$-phase sequence of length $n$ is equivalent to a so-called \emph{generalised bent function} on $\Z_n$, as studied extensively by Kumar, Scholtz, and Welch~\cite{KumSchWel1985}. In particular, some partial proofs for both directions of Conjecture~\ref{con:perfect_sequences} are given in~\cite{KumSchWel1985}. Most notably,~\cite[Property 6]{KumSchWel1985} shows that no perfect $n$-phase sequence of length~$n$ exists for $n\equiv 2\pmod 4$ when $2$ is semiprimitive modulo $n/2$ (see Section~\ref{sec:Nonexistence_perfect} for the definition of semiprimitivity). Mow~\cite{Mow1995} has proved the ``if'' direction of Conjecture~\ref{con:perfect_sequences}. Here we proceed in a way that is slightly different from Mow's treatment~\cite{Mow1995} and first show that it is sufficient to take $n$ to be a prime power. 
\par
Let $A$ and $B$ be sequences of length $n_1$ and $n_2$, respectively. We define $A\otimes B$ to be the sequence $C$ of length $n_1n_2$ defined by
\[
C(k)=A(k)B(k).
\]
Note that, if $A$ is an $H_1$-phase sequence and $B$ is an $H_2$-phase sequence, then $A\otimes B$ is an $H$-phase sequence, where $H=\lcm(H_1,H_2)$. The following result is an immediate consequence of the Chinese Remainder Theorem.
\begin{lemma}
\label{lem:product_periodic_corr}
Let $A$ and $B$ be sequences of length $n_1$ and $n_2$ with $\gcd(n_1,n_2)=1$. Then
\[
R_u(A\otimes B)=R_u(A)R_u(B)
\]
for every $u$. In particular, if $A$ and $B$ are perfect sequences, then $A\otimes B$ is a perfect sequence of length $n_1n_2$.
\end{lemma}
\par
We now consider perfect sequences whose length is a power of an integer (which is not necessarily prime). We distinguish the cases that the power is even or odd. For even powers, the length is a square, in which case we take a construction due to Heimiller~\cite{Hei1961} (who considered the special case where the length is $p^2$ for prime $p$) and Frank and Zadoff~\cite{FraZad1962}. For some reason, it is customary to call these sequences \emph{Frank} sequences.
\begin{definition}[Frank sequences]
\label{def:frank}
Let $m$ be a positive integer. A \emph{Frank sequence} $A$ of length $m^2$ is defined by
\[
A(j+km)=\exp\bigg(\frac{2\pi ijk}{m}\bigg),
\]
where $j$ and $k$ are integers satisfying $0\le j,k<m$.
\end{definition}
\par
We can think of a Frank sequence of length $m^2$ as the concatenation of the rows of the $m\times m$ matrix with entries $e^{2\pi ijk/m}$ at positions $(k,j)$.
\begin{theorem}[\cite{Hei1961}, \cite{FraZad1962}]
\label{thm:frank_perfect}
A Frank sequence of length $m^2$ is a perfect $m$-phase sequence.
\end{theorem}
\par
For lengths that are odd powers of an integer, we take a construction due to Milewski \cite{Mil1983}, building on earlier results due to Chu~\cite{Chu1972}.
\begin{definition}[Milewski and Chu sequences]
\label{def:milewski}
Let $m$ be a positive integer and let~$h$ be a nonnegative integer. A \emph{Milewski sequence} $A$ of length $m^{2h+1}$ is defined by
\[
A(j+km^h)=\begin{cases}
\exp\Big(\frac{\pi ik(2j+km^h)}{m^{h+1}}\Big)   & \text{for even $m$}\\[1.5ex]
\exp\Big(\frac{\pi ik(2j+(k+1)m^h)}{m^{h+1}}\Big) & \text{for odd $m$},
\end{cases}
\]
where $j$ and $k$ are integers satisfying $0\le j<m^h$ and $0\le k<m^{h+1}$. For $h=0$, we obtain a sequence of length $m$, which is also called a \emph{Chu sequence} of length $m$. 
\end{definition}
\par
We can think of a Milewski sequence of length $m^{2h+1}$ as the concatenation of the rows of the $m^{h+1}\times m^h$ matrix with entries
\[
C(k)\exp\bigg(\frac{2\pi ijk}{m^{h+1}}\bigg)
\]
at positions $(k,j)$, where $C$ is a Chu sequence of length $m$.
\begin{theorem}[\cite{Mil1983}]
\label{thm:milewski_perfect}
A Milewski sequence of length $n=m^{2h+1}$ is a perfect $H$-phase sequence, where
\[
H=\begin{cases}
2m      & \text{for $n\equiv 2\pmod 4$}\\
m^{h+1} & \text{otherwise}.
\end{cases}
\]
\end{theorem}
\par
As a consequence of Lemma~\ref{lem:product_periodic_corr} and Theorems~\ref{thm:frank_perfect} and~\ref{thm:milewski_perfect}, we obtain the following result, which proves the ``if'' part of Conjecture~\ref{con:perfect_sequences}.
\begin{theorem}
\label{thm:perfect_construction}
Let $n$ be an integer with $n>1$ and square-free part~$r$. Let
\[
n=p_1^{h_1}p_2^{h_2}\cdots p_s^{h_s}
\]
be the prime power factorisation of $n$. For each $k\in\{1,2,\dots,s\}$, let $A_k$ be either a Frank or a Milewski sequence of length $p_k^{h_k}$, depending on whether $h_k$ is even or odd, respectively. Then $A_1\otimes A_2\otimes\cdots\otimes A_s$ is a perfect $H$-phase sequence of length $n$, where
\[
H=\begin{cases}
2\sqrt{rn} & \text{for $n\equiv 2\pmod 4$}\\
\sqrt{rn}  & \text{otherwise}.
\end{cases}
\]
\end{theorem}
\par
Theorem~\ref{thm:perfect_construction} is essentially due to Mow~\cite{Mow1995}. We also refer to Mow~\cite{Mow1995},~\cite{Mow1996} and the references therein for more general constructions of perfect $H$-phase sequences.

\subsection{Aperiodic autocorrelation of nonbinary sequences}

Let $A$ be a sequence of length $n>1$. As in Section~\ref{sec:measures}, the \emph{peak sidelobe level} of $A$ is
\[
M(A)=\max_{0<u<n}\abs{C_u(A)}.
\]
Note that, if $A$ is a unimodular sequence, then $\abs{C_{n-1}(A)}=1$, and so $M(A)\ge 1$. Accordingly, following Golomb and Scholtz~\cite{GolSch1965}, we  define a \emph{unimodular} Barker sequence to be a unimodular sequence $A$ of length at least $2$ with $M(A)=1$. We also define an \emph{$H$-phase} Barker sequence analogously. Then a $2$-phase Barker sequence is a Barker sequence in the usual sense, discussed in Section~\ref{sec:barker}. It should be noted that it is possible to distinguish $H$-phase Barker sequences according to other measures for the collective smallness of the aperiodic autocorrelations (see Jedwab~\cite[\S~7]{Jed2008} for a detailed discussion).
\par
For fixed $H$, the existence problem of $H$-phase Barker sequences seems to be parallel to that for ordinary Barker sequences. While $H$-phase Barker sequences exist for small lengths (see~\cite[Table~1]{Jed2008} for $H\in\{2,3,4,6,8\}$), Jedwab~\cite{Jed2008} reports the nonexistence of $H$-phase Barker sequences of length $n$ for $H=3$ and $10\le n\le 76$, for $H=4$ and $16\le n\le 60$, for $H=6$ and $19\le n\le 29$, and for $H=8$ and $17\le n\le 25$. These data suggest the conjecture that, for fixed $H$, there are only finitely many $H$-phase Barker sequences.
\par
The situation seems to be completely different if we allow $H$ to grow with~$n$. Indeed, using the same heuristic reasoning as that leading to Conjecture~\ref{con:min_PSL}, Ein-Dor, Kanter, and Kinzel~\cite{EinKanKin2002} proposed a conjecture, which we summarise as follows.
\begin{conjecture}[\cite{EinKanKin2002}]
\label{con:min_PSL_n-phase}
Let $m_H(n)$ be the minimum peak sidelobe level over all $H$-phase sequences of length $n$. Then
\[
\lim_{n\to\infty}m_n(n)=1.
\]
\end{conjecture}
\par
On the other hand, using clever optimisation methods, the following is known (see Nunn and Coxson~\cite{NunCox2008a} for latest results).
\begin{proposition}
\label{pro:unimodular_Barker}
There exist unimodular Barker sequences for all lengths $n\le 70$ and for $n\in\{72,76,77\}$.
\end{proposition}
\par
It seems likely that Proposition~\ref{pro:unimodular_Barker} will be improved with the availability of more computing power. Indeed it seems plausible that unimodular Barker sequences exist for all lengths.
\begin{question}
Is there a unimodular Barker sequence for every length?
\end{question} 
\par
We now consider the aperiodic autocorrelations of specific families of unimodular sequences, namely Frank and Chu sequences (see Definitions~\ref{def:frank} and~\ref{def:milewski}). Turyn~\cite{Tur1967} calculated the peak sidelobe level of Frank sequences.
\begin{theorem}[\cite{Tur1967}]
\label{thm:psl_frank}
Let $A_n$ be a Frank sequence of length $n=m^2$. Then
\[
M(A_n)=\begin{cases}
1/\sin\big(\frac{\pi}{m}\big) & \text{for even $m$}\\[1ex]
2/\sin\big(\frac{\pi}{2m})   & \text{for odd $m$}.
\end{cases}
\]
In particular,
\[
\lim_{n\to\infty}\frac{M(A_n)}{n^{1/2}}=\frac{1}{\pi}=0.31830\dots.
\]
\end{theorem}
\par
Theorem~\ref{thm:psl_frank} shows that there exists an infinite family of unimodular sequences of length $n$ whose peak sidelobe level grows like a constant times $\sqrt{n}$. So far, this has not been proven for binary sequences (see Section~\ref{sec:PSL}).
\par
The asymptotic behaviour of the peak sidelobe level of Chu sequences is similar to that of Frank sequences, as shown by Mow and Li~\cite{MowLi1997}, although the leading constant is slightly larger.
\begin{theorem}[\cite{MowLi1997}]
Let $A_n$ be a Chu sequence of length $n$. Then
\[
\lim_{n\to\infty}\frac{M(A_n)}{n^{1/2}}=\frac{\sin\sigma}{\sqrt{\pi\sigma}}= 0.48026\dots,
\]
where $\sigma=1.16556\dots$ is the smallest positive root of $\tan x=2x$.
\end{theorem}
\par
Mow and Li~\cite{MowLi1997} also give an upper bound for the peak sidelobe level of Chu sequences of length $n$ that holds for each $n\ge 2$.
\par
We conclude this section with some results on the merit factor of families of unimodular sequences. Recall from Section~\ref{sec:merit_factor} that the \emph{merit factor} of a sequence $A$ of length~$n$ is
\[
F(A)=\frac{C_0(A)^2}{\displaystyle 2\sum_{0<u<n}\abs{C_u(A)}^2},
\]
provided that the denominator is nonzero.
\par
The merit factor of Chu sequences has been studied since at least 1961 by complex analysts, including Littlewood~\cite{Lit1961},~\cite{Lit1962},~\cite{Lit1966},~\cite{Lit1968} and Newman~\cite{New1965}. Independently, the problem was studied in the engineering literature~\cite{AntBom1990},~\cite{StaBoc2000},~\cite{Mer2013}. However the exact asymptotic behaviour of the merit factor of Chu sequences has only been established recently by the author~\cite{Sch2013}, correcting a false calculation by Littlewood~\cite{Lit1966}.
\begin{theorem}[\cite{Sch2013}]
\label{thm:mf_chu}
Let $A_n$ be a Chu sequence of length $n$. Then
\[
\lim_{n\to\infty}\frac{F(A_n)}{n^{1/2}}=\frac{\pi}{2}=1.57079\dots.
\]
\end{theorem}
\par
A stronger version of Theorem~\ref{thm:mf_chu} has been suggested previously by Borwein and Choi~\cite{BorCho2000}.
\begin{conjecture}[\cite{BorCho2000}]
Let $A_n$ be a Chu sequence of length $n$. Then
\[
\frac{n^{1/2}}{F(A_n)}=\frac{2}{\pi}+\frac{\delta_n}{3n}+O(n^{-2}),
\]
where $\delta_n=-2$ for $n\equiv 0,1\pmod 4$ and $\delta_n=1$ for $n\equiv 2,3\pmod 4$. 
\end{conjecture}
\par
The asymptotic behaviour of the merit factor of Frank sequences is also known, showing that Frank sequences are slightly better than Chu sequences with respect to the asymptotic merit factor.
\begin{theorem}[\cite{Sch2013}]
\label{thm:mf_frank}
Let $A_n$ be a Frank sequence of square length $n$. Then
\[
\lim_{n\to\infty}\frac{F(A_n)}{n^{1/2}}=\frac{\pi^2}{4}=2.46740\dots.
\]
\end{theorem}
\par
Theorem~\ref{thm:mf_chu} and~\ref{thm:mf_frank} show that the merit factor of unimodular sequences can grow without bound, which has not been proven so far for binary sequences.
\par
It should be noted that the asymptotic behaviour of the peak sidelobe level and the merit factor of general Milewski sequences is currently unknown.


\section{Golay pairs}
\label{sec:Golay}

\subsection{Definitions and a recursive construction}

In this section we study pairs of sequences $(A,B)$ of equal length $n$ with the property
\[
C_u(A)+C_u(B)=0\quad\text{for all $0<u<n$}.
\]
Such a pair is called a \emph{Golay (complementary) pair}. An nontrivial example of a Golay pair is given by the two sequences
\[
(1,1,1,-1)\quad\text{and}\quad(1,1,-1,1).
\]
Golay introduced these objects in 1951~\cite{Gol1951} for applications in spectrometry and studied them more  systematically in 1961~\cite{Gol1961}. Since then, Golay pairs have found many other applications, including coded aperture imaging~\cite{OhyHonTsu1978} (where Golay pairs have been rediscovered and called \emph{pinhole codes}), optical time domain reflectometry~\cite{Naz1989}, medical ultrasound~\cite{Now2003}, and multicarrier communications~\cite{Pop1991},~\cite{DavJed1999}. In particular, the latter application has revived the interest in Golay pairs in the last 15 years.
\par
The central question concerning Golay pairs is: For which lengths do Golay pairs consisting of sequences with entries from a given set exist? We shall study this problem for the most important cases of \emph{$H$-phase Golay pairs}, by which we mean that the two sequences in the pair are $H$-phase sequences. As usual, $2$-phase Golay pairs are also called \emph{binary} Golay pairs. For applications in multicarrier communications it is also important to know how many Golay pairs exist for a given length with entries from a given set, but we do not consider this problem here.
\par
It is surprisingly easy to construct Golay pairs, even \emph{binary} Golay pairs, for infinitely many lengths. Turyn~\cite[Lemma 5]{Tur1974} provided a simple recursive construction that produces a Golay pair of length $mn$ from two Golay pairs of length $m$ and~$n$.
\par
\begin{theorem}[{\cite{Tur1974}}]
\label{thm:Golay_recursive}
Let $\otimes$ be the Kronecker product and, for a sequence $A$, write $A^*$ for the sequence obtained by reading $A$ backwards. Let $(A,B)$ be a Golay pair of length $n$ and let $(X,Y)$ be a binary Golay pair of length $m$. Then the two sequences
\[
A\otimes \Big(\frac{X+Y}{2}\Big)+B\otimes \Big(\frac{X-Y}{2}\Big)\quad\text{and}\quad A\otimes \Big(\frac{X^*-Y^*}{2}\Big)-B\otimes \Big(\frac{X^*+Y^*}{2}\Big)
\]
form a Golay pair of length $mn$.
\end{theorem}
\par
The special case that $X=(1,1)$ and $Y=(1,-1)$ in Theorem~\ref{thm:Golay_recursive} was previously recognised by Golay~\cite{Gol1951},~\cite{Gol1961}. In this case, Theorem~\ref{thm:Golay_recursive} produces the Golay pair consisting of the sequences
\[
(A,B)\quad\text{and}\quad(A,-B).
\]
This is reminiscent of the recursion~\eqref{eqn:shapiro_recursion} that generates the Shapiro sequences. The Shapiro sequences can indeed be recovered by applying Theorem~\ref{thm:Golay_recursive} iteratively to $A=B=(1)$. It should also be noted that Theorem~\ref{thm:Golay_recursive} has several variations~\cite{FieJedPar2008}, all of which can be generalised to an array construction, which gives further Golay pairs (of the same lengths as those produced by Theorem~\ref{thm:Golay_recursive}) via a three-stage construction process~\cite{FieJedPar2008a}. 
\par
In the next two sections we summarise the knowledge on the existence question for binary and $H$-phase Golay pairs.

\subsection{Binary Golay pairs}

Binary Golay pairs are known for lengths $2$, $10$, and $26$, for example:
\begin{align*}
n&=2:&\begin{split}
(+\,+)\\
(+\,-)
\end{split}\\[1ex]
n&=10:&\begin{split}
&(++-+-+--+\,+)\\
&(++-+++++-\,-)
\end{split}\\[1ex]
n&=26:&\begin{split}
&(++++-++--+-+-+--+-+++--++\,+)\\
&(++++-++--+-+++++-+---++--\,-)
\end{split}
\end{align*}
Interestingly, as shown by Jedwab and Parker~\cite{JedPar2009}, these Golay pairs can be obtained from Barker sequences of odd length. By applying Theorem~\ref{thm:Golay_recursive} to these pairs, we obtain the following result.
\begin{corollary}
\label{cor:existence_binary_golay}
There exist binary Golay pairs for all lengths of the form $2^a10^b26^c$, where $a$, $b$, and $c$ are nonnegative integers.
\end{corollary}
\par
There is a particularly nice construction by Davis and Jedwab~\cite[Theorem 3]{DavJed1999} for binary Golay pairs of length a power of~$2$.
\begin{theorem}[{\cite{DavJed1999}}]
\label{thm:golay_power_of_two}
Let $m$ be a positive integer, let $\pi$ be a permutation of $\{1,2,\dots,m\}$, and let $e,e',e_1,\dots,e_m\in\{0,1\}$. Define the sequences $A$ and $B$ of length $2^m$ by
\begin{align*}
A(j_1+2j_2+\cdots+2^{m-1}j_m)&=(-1)^{\sum_{k=1}^{m-1}j_{\pi(k)}j_{\pi(k+1)}+\sum_{k=1}^me_kj_k+e}\\
B(j_1+2j_2+\cdots+2^{m-1}j_m)&=(-1)^{j_{\pi(1)}+e'}A(j_1+2j_2+\cdots+2^{m-1}j_m),
\end{align*}
where $j_1,\dots,j_m\in\{0,1\}$. Then $(A,B)$ is a binary Golay pair.
\end{theorem}
\par
In the case that $\pi$ sends $k$ to $m+1-k$ in Theorem~\ref{thm:golay_power_of_two}, we can again obtain the Shapiro sequences. Theorem~\ref{thm:golay_power_of_two} implies the following result~\cite[Corollary 5]{DavJed1999}.
\begin{corollary}[{\cite{DavJed1999}}]
There exist at least $2^{m+2}m!$ ordered binary Golay pairs of length~$2^m$.
\end{corollary}
\par
No binary Golay pair is known whose length is not of the form $2^a10^b26^c$ and no such Golay pair exists for lengths up to $100$, as shown by Borwein and Ferguson~\cite{BorFer2004} using clever exhaustive search methods. 
\begin{question}
\label{que:binary_golay}
Is there a binary Golay pair whose length is not of the form $2^a10^b26^c$ for some nonnegative integers $a,b,c$?
\end{question}
\par
It seems unlikely that Question~\ref{que:binary_golay} has a positive answer. Only two general results on the nonexistence of binary Golay pairs are known. The first is due to Golay himself~\cite{Gol1961} and the second is due to Eliahou, Kervaire, and Saffari~\cite{EliKerSaf1990}.
\begin{proposition}[\cite{Gol1961}]
\label{pro:Golay_even}
If there exists a binary Golay sequence pair of length $n>1$, then $n$ is even.
\end{proposition}
\par
\begin{proposition}[\cite{EliKerSaf1990}]
\label{pro:Golay_pmod4}
If there exists a binary Golay sequence pair of length $n>1$, then $n$ has no prime factor congruent to $3$ modulo $4$.
\end{proposition}
\par
A considerably simpler proof of Proposition~\ref{pro:Golay_pmod4} was later provided by Eliahou, Kervaire, and Saffari~\cite{EliKerSaf1991}. In fact,~\cite[Lemma 1.5]{EliKerSaf1991} contains the following more general result, from which Proposition~\ref{pro:Golay_pmod4} follows immediately.
\begin{theorem}[{\cite{EliKerSaf1991}}]
\label{thm:Golay_pmod4}
Let $p$ be an odd prime and suppose that $A,B\in\Z[z]$ are polynomials satisfying
\[
A(z)A(z^{-1})+B(z)B(z^{-1})\equiv 0\pmod p
\]
in $\Z[z,z^{-1}]$. Then $p$ is congruent to $1$ modulo $4$.
\end{theorem}

\subsection{Nonbinary Golay pairs}

We now summarise results on the existence pattern of $H$-phase Golay pairs. Since no two $H$-th roots of unity can sum to zero when $H$ is odd, we see that $H$-phase Golay pairs can only exist for even $H$. There exist $4$-phase Golay pairs of length $3$, $5$, $11$, and~$13$~\cite{Fra1980},~\cite{HolKar1994}. As shown by Gibson and Jedwab~\cite{GibJed20011}, these can  also be obtained from Barker sequences of odd length. It then follows from Theorem~\ref{thm:Golay_recursive} and Corollary~\ref{cor:existence_binary_golay} that there exist $4$-phase Golay pairs for all lengths of the form
\begin{equation}
\text{$2^a10^b26^cm$, where $a,b,c\ge 0$ are integers and $m\in\{1,3,5,11,13\}$}.   \label{eqn:lengths_Golay}
\end{equation}
No $H$-phase Golay pair is known whose length is not of the form~\eqref{eqn:lengths_Golay}, although there are $6$-phase Golay pairs of length a multiple of $10$~\cite{Fie2013} or a multiple of $16$~\cite{FieJedWie2010},~\cite{Fie2013} that are not (and cannot be trivially obtained from) binary Golay pairs. The smallest natural numbers not of the form~\eqref{eqn:lengths_Golay} are $7$, $9$, $14$, and $15$, and indeed, it has been verified with a computer~\cite{Fie2013} that there is no $H$-phase Golay pair of length $n$ for $n\in\{7,9\}$ and all $H\le 36$ and for $n\in\{14,15\}$ and all $H\le 10$. 
\begin{question}
Is there an $H$-phase Golay pair whose length is not of the form~\eqref{eqn:lengths_Golay}? In particular, is there an $H$-phase Golay pair of odd length $n>13$?
\end{question}
\par
From Proposition~\ref{pro:Golay_even} we know that there is no binary Golay pair of odd length $n>1$. Fiedler~\cite[Conjecture 1]{Fie2013} conjectured the more general assertion that there is no $H$-phase Golay pair of odd length $n>1$ whenever $H\equiv 2\pmod 4$.
\begin{conjecture}[{\cite{Fie2013}}]
\label{con:Hn_2mod4}
If $H\equiv 2\pmod 4$, then there is no $H$-phase Golay pair of odd length $n>1$.
\end{conjecture}
\par
Conjecture~\ref{con:Hn_2mod4} holds for values of $H$ having a small odd prime divisor, which can be deduced from the following result~\cite[Corollary~5.2]{Fie2013}.
\begin{proposition}[{\cite{Fie2013}}]
Let $H>2$ be an integer satisfying $H\equiv 2\pmod 4$ and let~$p$ be the smallest odd prime factor of $H$. If there exists an $H$-phase Golay pair of odd length $n$, then $n<2p$.
\end{proposition}
\par
For example, there is no $6$-phase Golay pair of odd length greater than $6$. The nonexistence of $6$-phase Golay pairs of length $3$ and $5$ is easily established, so the assertion of Conjecture~\ref{con:Hn_2mod4} is true for $H=6$. Similarly, the assertion of Conjecture~\ref{con:Hn_2mod4} is true for all $H\le 36$~\cite{Fie2013}.
\par
Fiedler~\cite{Fie2013} also proved the following result, which crucially relies on Theorem~\ref{thm:Golay_pmod4} and complements Proposition~\ref{pro:Golay_pmod4}. 
\begin{proposition}[{\cite{Fie2013}}]
Let $p$ be an odd prime congruent to $3$ modulo $4$ and let $H$ be twice a power of $p$. If there exists an $H$-phase Golay pair of length $n$, then~$p$ does not divide~$n$.
\end{proposition}
\par
For example, there are no $6$-phase Golay pairs of lengths $3$, $6$, $9$, $12$, \dots.


\section*{Acknowledgements}

I would like to thank Christian G\"unther, Jonathan Jedwab, Dieter Jungnickel, and Peter Wild for some careful comments on a draft of this survey.



\end{document}